\documentclass[%
 reprint,
 superscriptaddress,
 amsmath,amssymb,
 aps,
]{revtex4-2}

\usepackage{graphicx}
\usepackage{dcolumn}
\usepackage{bm}

\usepackage{color} 
\usepackage{xcolor}
\begin{document}

\preprint{APS/123-QED}

\title{Photoionization of $nS$ and $nD$ Rydberg atoms of Rb and Cs from the near-infrared to the ultraviolet spectral region}

\author{Michael A. Viray}
\email{mviray@umich.edu}
\affiliation{Department of Physics, University of Michigan, Ann Arbor, Michigan, 48109, USA}

\author{Eric Paradis}
\affiliation{Department of Physics and Astronomy, Eastern Michigan University, Ypsilanti, Michigan, 48197, USA}%

\author{Georg Raithel}
\affiliation{Department of Physics, University of Michigan, Ann Arbor, Michigan, 48109, USA}

\date{\today}

\begin{abstract}
We present calculations of the photoionization (PI) cross sections of rubidium and cesium Rydberg atoms for light with wavelengths ranging from the infrared to the ultraviolet, using model potentials from [M. Marinescu, H. R. Sadeghpour, and A. Dalgarno, Phys. Rev. {\bf{A}} 49, 982 (1994)]. The origins of pronounced PI minima are identified by investigating the free-electron wavefunctions. These include broad PI minima in the $nS$ to $\epsilon P$ PI channels of both Rb and Cs, with free-electron energy $\epsilon$, which are identified as Cooper minima. Much narrower PI minima in the $nD$ to $\epsilon F$ channels are due to shape resonances of the free-electron states.   
We describe possible experimental procedures for measuring the PI minima, and we discuss their implications in fundamental atomic physics as well as in practical applications.
\end{abstract}

\maketitle


\section{Introduction}

Photoionization (PI), or the photoelectric effect, of atoms is one of the longest-studied processes in atomic physics, dating back to at least the Bohr model~\cite{bohr1913}. Because the outermost electron(s) of Rydberg atoms lie at high energy levels, the minimum photon energy required for Rydberg-atom PI typically is in the milli-eV regime. Hence, the spectral range of the ionizing radiation 
ranges from THz fields through the infrared (IR), visible and ultraviolet (UV) spectral regions and higher. Rydberg atoms are known to be extremely sensitive to PI by black-body radiation, a consequence of their extended electronic wavefunctions and large transition electric-dipole moments~\cite{gallagher, friedrich}. Generally, as the photon energy increases, the Rydberg atoms becomes less sensitive to PI due to increasing mismatch between bound-state and free-electron wavefunctions, and the ionization cross sections drop rapidly. However, in certain PI channels this general trend is 
interrupted by pronounced minima in the 
PI cross sections as a function of wavelength $\lambda$ of the ionizing field. The PI minima are quite sensitive to the assumed Rydberg-electron model potentials. Hence, experimental 
PI studies can serve as a method to test and to possibly fine-tune model potentials. 
Data on Rydberg-atom PI cross sections are also relevant to applications of Rydberg atoms in which the rate of laser-induced PI must be minimized.


PI minima in atoms can be attributed to Cooper minima or shape resonances, which are, in the following, briefly explained. Cooper minima, first reported by John Cooper in 1962~\cite{cooper62}, occur as a result of vanishing PI matrix-element integrals.
The PI matrix elements are integrals over expressions that involve a product of an initial- and a final-state electron wavefunction. The wavefunctions are quasi-periodic and have different spatial periods and phases. At certain values of $\lambda$, the matrix-element integral can vanish, in remote resemblance to destructive interference between two out-of-phase periodic functions. The free-state wavefunction on its own does not exhibit any special behavior at the Cooper minima. The associated dips in the PI cross sections as a function of $\lambda$ can be hundreds of nm wide.
Shape resonances, on the other hand, are due to quasi-bound scattering states within an inner well of the relevant free-electron potential. As in our case, the potential barrier that separates the outer region from the inner well can
arise from the sum of a short-range attractive potential with a repulsive core and the long-range centrifugal potential, $\hbar^2 \ell (\ell +1) / (2m_e r^2)$, with electron angular momentum $\ell$, electron mass $m_e$, and electron radial coordinate $r$. As $\lambda$ is varied, the free-electron wavefunction can pass through a narrow resonance in the inner well, characterized by a large-amplitude quasi-bound scattering state inside the barrier and a $\pi$ phase shift outside the barrier. In certain Rydberg-atom PI channels, shape resonances cause minima in the PI cross sections; these tend to be narrower as a function of $\lambda$ than Cooper minima.


Finding the wavelengths at which PI cross sections have minima is interesting from a basic atomic-physics perspective, because it represents a test of the assumed model potentials for the Rydberg and the free-electron states. Knowledge of PI minima can also be beneficial from an applications standpoint. Rydberg atoms in optical dipole traps and optical lattices can become photo-ionized by the trapping beams themselves. If the trapping- or lattice-beam wavelength is set at a PI minimum, the trapped atoms will be less prone to PI by the trapping beams. The absence of Rydberg-atom PI could be important, for instance, in experiments on quantum simulation and quantum information processing using Rydberg atoms~\cite{Saffman.2005a, Zhang.2011, Nguyen.2018, Barredo.2020}, quantum control~\cite{Lampen.2018, Cardman.2020a}, high-precision spectroscopy~\cite{Moore.2015a, Moore.2015b, Ramos.2017, Malinovsky.2020a}, and in large-scale arrays of Rydberg atoms that could potentially be trapped~\cite{bernien17}.

Various experiments and theoretical investigations have been performed over the years to find PI minima, both from Cooper minima and from shape resonances. Regarding Cooper minima, Zatsarinny and Tayal computationally calculated Cooper minima in potassium \cite{zatsarinny10}, which were later verified experimentally by Yar, Ali, and Baig \cite{yar13}. Beterov et. al. calculated transition probabilities and PI cross sections of alkali metals using a Coulomb approximation and a quasiclassical model \cite{beterov12}. Additionally, there have been measurements of the Cooper minimum of atoms in a plasma background \cite{sahoo06, lin10, lin11}. Meanwhile, shape resonances have been observed in positronium \cite{kar12} and in atom collisions \cite{boesten96, boesten97, yao19}, and they have been used to form Rydberg molecules~\cite{hamilton02, peper20}.

In this work, we report computational findings on Cooper minima and shape resonances in Rydberg-state PI in Rb and Cs. We use model potentials for rubidium and cesium from Refs.~\cite{marinescu94_dispersion} to determine the initial (bound) and and final (free-state) wave functions of the photoionized electrons. We compute PI cross sections across a wide range of $\lambda$, and find several Cooper minima and shape resonances. Results are evaluated in context with free-state wavefunction plots and the underlying model potentials, and comparisons between Cs and Rb are made. We discuss the viability of measuring these PI minima experimentally, as well as the relevance to atomic physics theory and to applications.

\section{Theory Background}
\label{sec:theory}

\subsection{Atomic model potentials}
\label{subsec:theory_potentials}

We denote the initial Rydberg states $\vert i \rangle = \vert n, \ell, m_\ell \rangle$ with principal, angular-momentum and magnetic 
quantum numbers $n, \ell$ and $m_\ell$, respectively, and the photo-ionized
free-electron states $\vert f \rangle = \vert \epsilon', \ell', m'_\ell \rangle$ with free-electron energy $\epsilon'$, and angular-momentum and magnetic quantum numbers $\ell'$ and $m'_\ell$, respectively.

The calculation of PI cross sections requires a procedure to calculate the initial-state and free-state wavefunctions, $\psi_i({\bf{r}}) = \langle {\bf{r}} \vert i \rangle$ and  $\psi_f({\bf{r}}) = \langle {\bf{r}} \vert f \rangle$, with relative electron position ${\bf{r}}$. The fine structure is neglected in the present work because it is much smaller than the Rydberg-atom binding energy and the energy of the free electron. 
The wavefunction calculation requires a set of atomic model potentials. Here,  we use model potentials for Rb and Cs developed and employed in Refs.~\cite{marinescu94_dispersion, marinescu94_2photon,  
marinescu94_dynamic}.
The model potentials $V_{0, \ell}(r)$ include correction terms to the Coulomb potential that yield the correct core-penetration and ion-core polarization quantum defects of the atomic energy levels for various angular momenta $\ell$. The model potentials depend on atomic species and on $\ell$, with the potentials of any one species being the same for all $\ell \geq 3$~\cite{marinescu94_dispersion}.
Including the (species-independent) centrifugal term yields effective potentials 
\begin{equation} \label{eq:pots}
    V_{\ell}(r) = V_{0,\ell}(r) + \frac{\hbar^2 \ell (\ell + 1)}{2 m_e r^2} \quad.
\end{equation}
The wavefunctions are calculated using these potentials with a numerical method outlined by Reinhard et. al. \cite{reinhard07}.

\subsection{Calculating PI Cross Sections}
\label{subsec:theory_cross}

PI is an effect of the $\hat{\bf{A}} \cdot \hat{\bf{p}}$-interaction of the minimal-coupling Hamiltonian~\cite{friedrich} in first order. Given a linearly polarized plane wave with polarization unit vector $\hat{\mathbf{n}}$, wave vector $\mathbf{k}$ and angular frequency $\omega$, the partial PI cross section is 
\begin{equation}
    \sigma = \frac{\pi e^2 \hbar^2}{\varepsilon_0 m_e^2 \omega c} \left| \hat{\mathbf{n}} \cdot \int \psi_f^* e^{i \mathbf{k} \cdot \mathbf{r}} \nabla \psi_i d^3r \right|^2 \left( \frac{1}{E_H a_0^2} \right) \, ,
\end{equation}
with the atomic energy unit $E_H \approx 27.2$~eV, and standard identifiers for other physical constants. The result is in SI units, m$^2$. The matrix element is computed in atomic units, with free states normalized in units of energy, 
i.e. $\langle \epsilon', \ell', m'_\ell \vert \epsilon'', \ell'', m''_\ell \rangle = \delta(\epsilon'-\epsilon'') \delta_{\ell'', \ell'}  \delta_{m''_\ell, m'_\ell}$, and the term in parentheses within Equation~2 converting the matrix-element square from atomic into SI units. 

It is shown elsewhere that for light-induced PI of Rydberg atoms the electric-dipole approximation (EDA), $e^{i \mathbf{k} \cdot \mathbf{r}}=1$, is valid at a level better than $10^{-4}$, {\sl{ i. e.}} electric-dipole-forbidden transitions have cross sections that are smaller than those of the dipole-allowed ones by a factor of at least ten thousand. We make the EDA and average the cross sections over the initial-state magnetic quantum number, $m_\ell$, to obtain the shell-averaged partial cross section,
\begin{equation}\label{sigma_av}
\bar{\sigma}_{n, \ell}^{\epsilon', \ell'}=\frac{\pi e^{2} \hbar^2 }{3\epsilon_{\rm 0} m_{\rm{e}}^2 \omega c}\frac{\ell_{>}}{(2\ell+1)}|M|^{2}\left(\frac{1}{E_{H} \it a_{\rm 0}^{\rm 2}}\right),
\end{equation}
\noindent where $M$ is a radial matrix element in atomic units,
\begin{equation}\label{sigma_av2}
M = \int_{0}^\infty u_{\epsilon',\ell'}(r)\left[
u'_{n,\ell}(r) \mp \frac{u_{n,\ell}(r)}{r}\ell_{>} \right] dr.
\end{equation} 
There, the upper sign is for $\ell_> = \ell' = \ell+1$ and the bottom sign for $\ell_> = \ell = \ell'+1$. The functions $u_{*,\ell}(r)$ are given by $u_{*,\ell}(r) = r R_{*,\ell}(r)$, with the usual radial wavefunction $R_{*,\ell}(r)$, and $* = n$ and $* = \epsilon'$ denoting the principal quantum number of the bound- and the energy of the free-electron state, respectively. The free-state energy follows from the wavelength of the PI light, $\lambda$, and the binding energy of the Rydberg atom. The free-state energy in atomic units is $\epsilon' = 2 \pi a_0 / (\alpha \lambda) - 1/(2n^{*2})$, where $\lambda$ is entered in meters, the fine structure constant $\alpha$, the effective quantum number $n^* = n - \delta_{n, \ell}$, and the quantum defect $\delta_{n, \ell}$.

For $z$-polarized light, $m_\ell$ is conserved, and the $m_\ell$-dependent PI cross sections follow from the shell-averaged ones via
\begin{equation}\label{sigmazm}
\sigma_{z,n,\ell,m_\ell}^{\epsilon',\ell'}
=\frac{3(\ell_{>}^{2}-m_\ell^{2})}{(2\ell_{>}+1)(2\ell_{>}-1)}\frac{(2\ell+1)}{\ell_{>}}
\bar{\sigma}_{n, \ell}^{\epsilon', \ell'} \quad ,
\end{equation}
with similar expressions applicable to other light polarizations. Therefore, it is sufficient to discuss the PI behavior in terms of the shell-averaged 
$\bar{\sigma}_{n, \ell}^{\epsilon', \ell'}$.

It is noted that the matrix element in Eq.~\ref{sigma_av2}
follows directly from the $\hat{\bf{A}} \cdot \hat{\bf{p}}$-interaction and the EDA. The matrix-element form in Eq.~\ref{sigma_av2} is known as velocity form~\cite{bethe, friedrich}. In the case that the atomic potential is velocity-independent (not including the centrifugal term), the matrix elements can be converted into length form~\cite{bethe, friedrich}, allowing an alternate, commonly used method to compute electric-dipole matrix elements. If the potentials are $\ell$-dependent, the velocity form must be used. We did confirm in our work that the PI cross sections that follow from matrix elements calculated in the velocity and in the length forms are identical for $\ell \geq 4$. This is expected, because for $\ell \geq 4$ only a single, $\ell$-independent model potential applies (namely, $V_{0,\ell \ge 3}= V_{0,\ell = 3}$) to compute {\sl{both}} the bound- and free-state wavefunctions.  As $\ell$ approaches zero, deviations between velocity- and length-form cross sections increase and reach about 15$\%$ relative difference for $\ell = 0$ (away from Cooper minima). Also, the exact $\lambda$-values at which the Cooper minima and shape resonances occur are slightly different. These deviations are due to the fact that the length form becomes increasingly inaccurate when $\ell$ approaches 0, because the
model potentials become increasingly $\ell$-dependent, whereas the velocity form remains accurate. For the PI cross-section calculations presented in this paper, we have chosen $\ell = 0$ and  $\ell = 2$, because these are the  experimentally most relevant cases. For these
$\ell$-values it is important to use the expressions given in Eqs.~\ref{sigma_av} and~\ref{sigma_av2}, which are in velocity form. 

We note that the difference between electric-dipole matrix elements
calculated in length and velocity forms is orders of magnitude lower for bound-bound transitions between Rydberg states than it is for Rydberg-atom PI. 
For bound-bound Rydberg transitions, which are in the microwave to THz range, the length-form
matrix elements are accurate enough for most purposes. However, the length form is not generally applicable to bound-free transitions, particularly in the optical regime. This finding can be explained, qualitatively, through the fact that the lower the transition energy is 
the more the matrix-element integral 
becomes dominated by contributions from the outer reaches of the electron configuration space, where all model potentials converge into an $\ell$-independent Coulomb potential.


\section{Rubidium PI Cross-Sections}
\label{sec:Rb}

\begin{figure*}
    \centering
    \includegraphics[width = 16 cm]{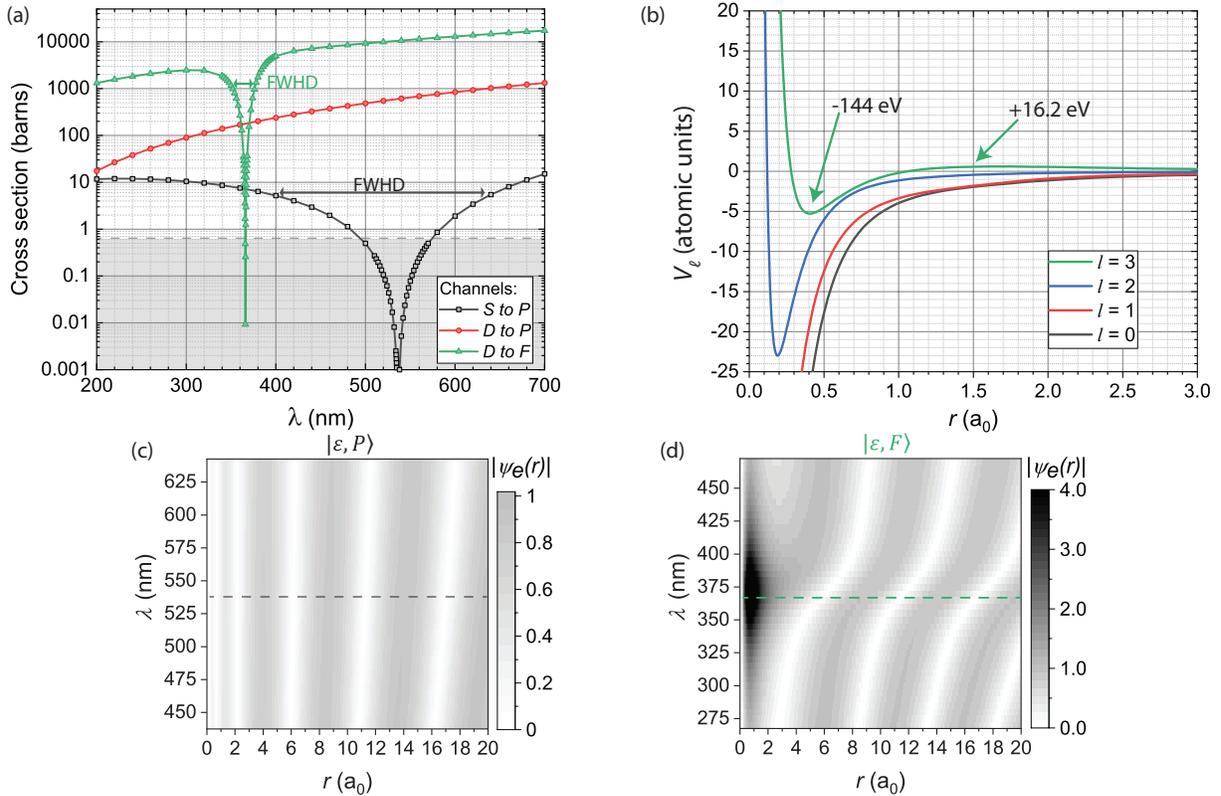}
    \caption{Rubidium: (a) Partial PI cross sections of $35S$ and $35D$ Rb Rydberg atoms vs PI wavelength $\lambda$: $S \rightarrow P$ (black squares), $D \rightarrow P$ (red circles), and $D \rightarrow F$ (green triangles). The dashed line shows the 
    Thomson scattering cross section ($\sigma_T = 0.665$~barn), which is the elastic photon scattering cross section of the Rydberg atoms. In the gray region, elastic scattering exceeds PI. 
    (b) Core region of the potentials $V_\ell(r)$ for $\ell = 0, 1, 2, 3$ in atomic units. (c) $P$-state free-electron wave function vs $\lambda$, with the
    Cooper minimum of the PI cross section indicated as a dashed line. The free-electron wavefunction exhibits an oscillatory pattern with a smooth, gradual phase shift as a function of $\lambda$, with 
    no marked behavior at the Cooper minimum. (d) 
    $D$-state free-electron wave function vs $\lambda$, with the 
    shape resonance in the $D \rightarrow F$ PI cross section indicated as a dashed line. The plot shows a quasi-bound state centered at the shape resonance.}
    \label{fig:Rb_plots}
\end{figure*}

Figure~\ref{fig:Rb_plots}(a) shows shell-averaged cross sections of
three PI channels of Rb Rydberg atoms, namely
$\bar{\sigma}_{n=35, \ell=0}^{\epsilon', \ell'=1} (\lambda)$, $\bar{\sigma}_{n=35, \ell=2}^{\epsilon', \ell'=1} (\lambda)$, and $\bar{\sigma}_{n=35, \ell=2}^{\epsilon', \ell'=3} (\lambda)$, for wavelengths ranging from the deep UV to the near-IR regime.
The free-electron energy in atomic units is $\epsilon' = 2 \pi a_0 / (\alpha \lambda) - 1/(2n^{*2})$. Aside from the resonant features discussed later, it is seen that the $S \rightarrow P$ cross sections are fairly small over the entire range, topping out at only about 20 times the Thomson cross section, $\sigma_T=0.665$~barn. In the near-IR spectral range, $nS$-type Rydberg atoms have PI cross sections that are lower than those of other low-$\ell$ PI channels by up to about three orders of magnitude. In contrast, the $D \rightarrow P$ and $D \rightarrow F$ channels in Fig.~\ref{fig:Rb_plots}(a) follow the generic trend that PI rates rapidly increase at longer wavelengths. Between $\lambda = 200$~nm and 700~nm, the PI cross sections of those channels increase by one to two orders of magnitude. Typically, the PI threshold will peak at the PI threshold, with depends on $n$ and $\ell$ and is in the far-IR regime.

The plot $\bar{\sigma}_{n=35, \ell=0}^{\epsilon', \ell'=1} (\lambda)$  in Fig.~\ref{fig:Rb_plots}(a) shows a minimum centered at $\lambda = 536$~nm. As the
$S \rightarrow P$ channel is the only PI channel of $S$-type Rydberg atoms, the total PI cross section equals the partial cross section $\bar{\sigma}_{n=35, \ell=0}^{\epsilon', \ell'=1} (\lambda)$, and is below $\sigma_T$ over a range 490~nm $\lesssim \lambda \lesssim$ 570~nm. 
Taking into account the asymmetry of the PI minimum in $\bar{\sigma}_{n=35, \ell=0}^{\epsilon', \ell'=1}(\lambda)$, we define the full width at half depth (FWHD) of the minimum as the range over which the cross section dips below half of the PI maximum seen in the UV range. For the Rb Cooper minimum, the PI maximum in the UV range is 10~barns, so the FWHD is the range over which the cross section dips below 5~barns. The FWHD is $\approx 240$~nm, with the FWHD-range covering the spectral region $400 \lesssim \lambda \lesssim 640$~nm.  The large width of the PI minimum
in the $S \rightarrow P$ channel serves as an indicator that this is a Cooper minimum, as will be proven below.

The $D \rightarrow F$ channel shows a minimum centered at $\lambda = 366$~nm with a FWHD of only about 10~nm. The narrow width of this minimum is a first indicator that this minimum 
is different in nature from the minimum in the
$S \rightarrow P$ channel; below we will show that the minimum in the $D \rightarrow F$ channel is due to a shape resonance. The $D \rightarrow P$ channel has no PI minimum within the range displayed in Fig.~\ref{fig:Rb_plots}(a). We also did not find any minimum over an extended search across a wider range from 100~nm to 2~$\mu$m (not shown). The total shell-averaged PI cross section of Rb $35D$ is given by the sum of the partial cross sections, $\bar{\sigma}_{n=35, \ell=2}^{\epsilon', \ell'=1} (\lambda)$ $+$ $\bar{\sigma}_{n=35,\ell=2}^{\epsilon', \ell'=3} (\lambda)$. Hence, the observable cross section of the $35D$-state is the sum of two of the curves in Fig.~\ref{fig:Rb_plots}(a).
The total 35$D$ PI cross section therefore has a minimum of about 200~barns at $366$~nm, and rises to about 3000~barns several tens of nm away.

The physical differences between the PI minima in the Rb $S \rightarrow P$ and $D \rightarrow F$ channels become apparent when looking at the inner regions of the relevant free-electron model potentials, $V_{\ell}(r)$, and the free-electron wavefunctions.  In Fig.~\ref{fig:Rb_plots}(b) we show $V_{\ell}(r)$ for 
$\ell = 0$ to 3 and over the range $r \leq 3~a_0$, and in Figs.~\ref{fig:Rb_plots}(c) and~(d) free-electron wave-function moduli for the $S \rightarrow P$ and $D \rightarrow F$ ionization channels, respectively, over the range $r \leq 20~a_0$. Figs.~\ref{fig:Rb_plots}(b)-(d) are focused on the central atomic region, where phase shifts and shape resonances determine the PI behavior. In Fig.~\ref{fig:Rb_plots}(c), the free-electron wavefunction does not exhibit any noteworthy feature, as $\lambda$ passes through the PI minimum. There merely is a gradual phase shift of the wavefunction due the slightly changing de-Broglie wavelength of the free electron. Hence, the PI minimum is due to incidental near-perfect destructive interference between bound-state and free-electron wavefunctions (where 
the bound-state function shows up in a modified form; see Eq.~\ref{sigma_av2}). This fact makes the PI minimum in the 
Rb $S \rightarrow P$ channel a Cooper minimum. Due to the rather gradual change of the free-state wavefunction, the Cooper minimum is comparatively wide in PI wavelength and free-electron energy. The Cooper minimum relates to the fact that the bound-state and free-state quantum defects differ by about 1/2 in Rb (the quantum defects are near 3.13 for $S$ and  2.65 for $P$).

The free-state wavefunction of the $D \rightarrow F$ channel exhibits a resonant structure at the PI minimum, which manifests in the region of high wavefunction amplitude near 
$r \approx 1 a_0$, as well as a phase shift by $\approx \pi$ in the outer, oscillatory region of the wavefunction. The resonance occurs at the $\lambda$-value of the PI minimum,
and it has a width in $\lambda$ of only a few tens of nm. Another piece of insight follows from the $V_{\ell}(r)$, plotted in Fig.~\ref{fig:Rb_plots}(b). There, it is seen that $V_{\ell=3}(r)$ exhibits an inner well formed by the centrifugal potential and the core region of the model potential $V_{0, \ell=3}(r)$ for $r \lesssim 1.5 a_0$. (Note Eq.\ref{eq:pots} regarding the definition of  $V_{0, \ell=3}$ and  $V_{\ell=3}$).  Of the four potentials shown, the $\ell = 3$ potential is the only one with 
an inner potential well and a centrifugal barrier.
The potential barrier peaks at an energy of about $+16.2$~eV, and the inner well reaches a minimal value of about $-144$~eV. It is apparent that the resonance, which is at about $\epsilon'=3.37$~eV, is the lowest and only quasi-bound state in the inner potential 
well of $V_{\ell=3}(r)$. These findings sum up to the statement that the PI minimum in 
the $D \rightarrow F$ channel is due to a shape resonance.
The phase shift of $\pi$ of the scattering wavefunction
across the resonance, seen in 
Fig.~\ref{fig:Rb_plots}(c), is another telltale sign of a shape resonance \cite{sakurai}.

After identifying the physical origins of the PI minima, we wish to comment on a length-form calculation of the PI cross sections, which has yielded qualitatively similar behaviors with quantitatively notable differences (not shown). For instance, the Cooper minimum in the length-form result for Rb $\bar{\sigma}_{n=35, \ell=0}^{\epsilon', \ell'=1} (\lambda)$ is shifted up in wavelength by about 50~nm relative to the velocity-form result. While the (unphysical) shift in the length-form result is less than the width of the minimum, it is large enough to be of significance. The shift represents a case in which the length-form calculation exhibits a significant error relative to the velocity-form calculation. We reiterate here that the velocity-form calculation is correct, while the length-form
is expected to show an error because of the $\ell$-dependence of the model potentials $V_{0, \ell}$. For $\ell \geq 4$ the PI cross sections obtained with the two forms agree, because the model potentials to be used for the bound ($\ell$) and free-electron ($\ell'=\ell \pm 1$) states are identical (namely,  $V_{0, \ell = 3}$).

It is also worth commenting on free-electron photon scattering due to the $A^2$-term in the atom-field interaction~\cite{friedrich}. Since the $\lambda$-range discussed in our work exceeds the atomic energy scale by orders of magnitude, the photon scattering has a cross section given by the Thomson scattering cross section, $\sigma_T=0.665$~barn.
For a truly free electron, the Compton (recoil) energy would be in the range of $h \times 10$~GHz.
This recoil energy would be too small to cause atomic bound-bound transitions, nor does it cause photo-ionization of the atom (as in typical instances of the Compton effect). Therefore, the Thomson scattering of a Rydberg electron is perfectly recoil-free and elastic (except for a very small recoil of the entire atom). In the gray regions in Figs.~\ref{fig:Rb_plots} and~\ref{fig:Cs_plots}, the elastic (Thomson) scattering rate exceeds the PI rate. 
Comparing the two effects, we further note that the Thomson scattering is due to the $A^2$-term in the atom-field interaction, and it occurs in the outer reaches of the Rydberg atom, $r \gtrsim 10~a_0$, where the Rydberg electron resides with near-unity probability, whereas 
PI (photo-electric effect) is due to the ${\bf{A}} \cdot {\bf{p}}$-term and occurs in the atomic core, $r \lesssim 10~a_0$.

The dependence of elastic photon scattering and PI on principal quantum number $n$ also is of interest. 
The elastic cross section, $\sigma_T$, is independent of $n$. In contrast, the PI cross sections, away from the resonances, have a generic scaling close to $\propto n^{*-3}$, with $n^* = n - \delta_{\ell}$ denoting the effective quantum number, and $\delta_{\ell}$ the leading term of the quantum defect. We have verified this scaling in additional calculations (which are not presented in  
detail).

\section{Cesium PI Cross-Sections}

\begin{figure*}
    \centering
    \includegraphics[width = 16 cm]{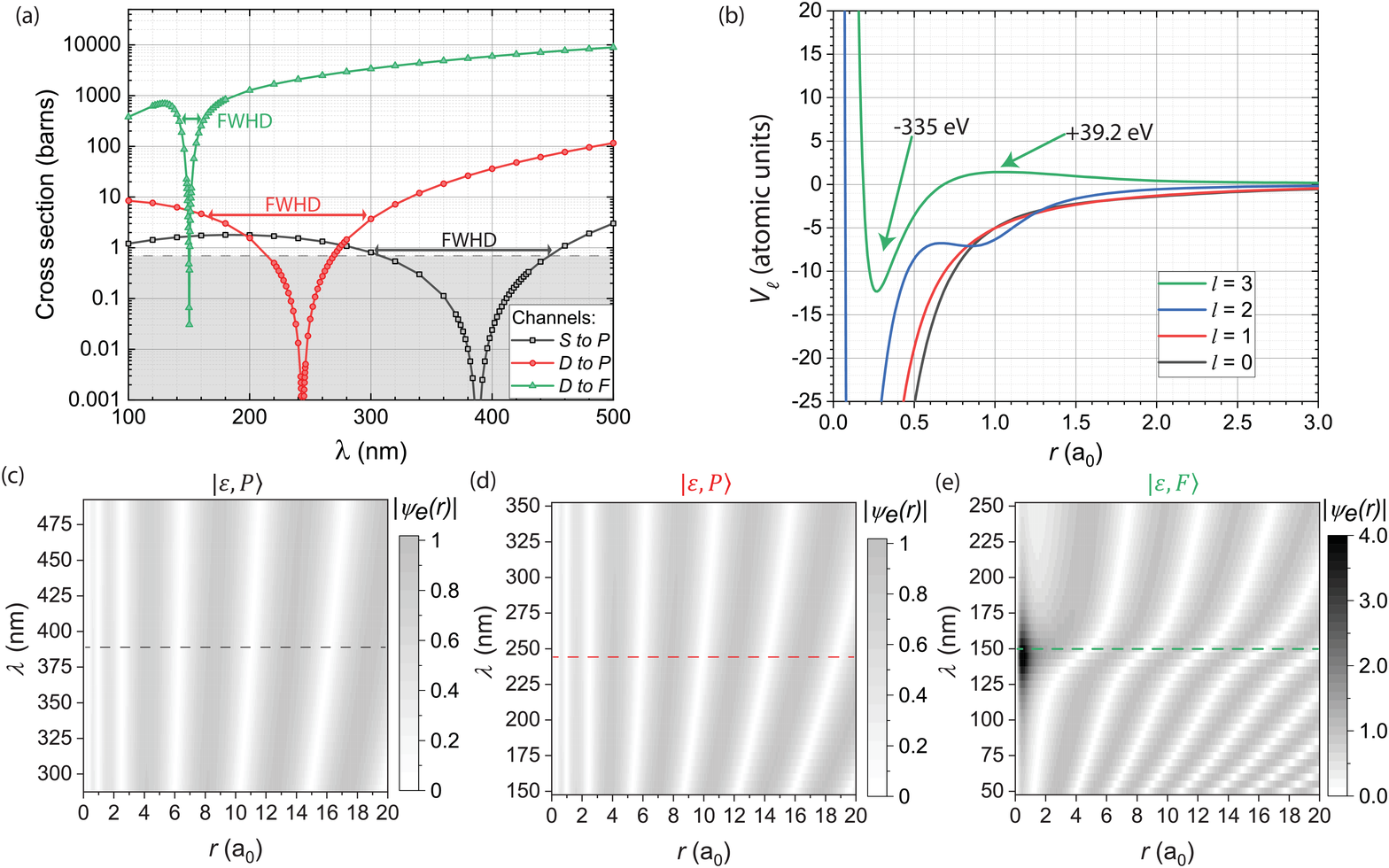}
    \caption{Cesium: (a) Partial PI cross sections of $35S$ and $35D$ Cs Rydberg atoms vs PI wavelength $\lambda$: $S \rightarrow P$ (black squares), $D \rightarrow P$ (red circles), and $D \rightarrow F$ (green triangles). The Thomson scattering cross section, $\sigma_T$, is indicated as in Fig.~\ref{fig:Rb_plots}.
    (b) Potential curves $V_{\ell}(r)$ for $\ell = 0, 1, 2, 3$ in atomic units.
    (c) $P$-state free-electron wave function vs $\lambda$, with the
    $S \rightarrow P$ Cooper minimum of the PI cross section indicated as a dashed line.
    (d) Same as (c), with the $D \rightarrow P$ Cooper minimum of the PI cross section indicated as a dashed line.
    (e) $D$-state free-electron wave function vs $\lambda$, with the shape resonance in the $D \rightarrow F$ PI cross section indicated as a dashed line. }
    \label{fig:Cs_plots}
\end{figure*}

Figure~\ref{fig:Cs_plots}~(a) shows partial PI cross sections 
for Cs $35S$ and $35D$, for the $S \rightarrow P$, $D \rightarrow P$, and $D \rightarrow F$ ionization channels. In this figure, which is organized analogous to Fig.~\ref{fig:Rb_plots}, we see that all three partial PI cross section channels have minima, namely Cooper minima for $S \rightarrow P$ and $D \rightarrow P$, and a shape resonance for $D \rightarrow F$. 
The PI minima are 100 to 200~nm deeper in the UV than in Rb. The PI cross section of the $35S$ state barely rises above the elastic scattering cross section, $\sigma_T$, across the displayed range, making $S$-type Rydberg atoms of Cs essentially PI-free at all wavelengths shorter than about 500~nm. Otherwise the trends observed 
in Fig.~\ref{fig:Cs_plots}~(a) follow those of Rb.  
As in Rb, in Cs the shape resonance is considerably narrower than the Cooper minima. The Cooper minimum in the $D \rightarrow P$ channel is of little relevance, because in the total PI cross section it will be near-invisible against PI on the $D \rightarrow F$ channel.

The potential curves and free-state wavefunction maps for Cs, shown in Fig~\ref{fig:Cs_plots}~(b) 
and Fig~\ref{fig:Cs_plots}~(c-e), respectively, present a situation that is similar to that in Rb. The Cooper minima in the $S \rightarrow P$ and $D \rightarrow P$ channels have a FWHD of about 100~nm and are characterized by free-state wavefunctions with smooth, $\lambda$-dependent phase changes and without any resonant behavior. 
The $\ell = 3$ potential is the only one that features a relevant barrier, which is located at $r \approx 1~a_0$ and peaks at 39.2~eV. The potential well inside the barrier bottoms out at $-335$~eV. The lowest electron ``state'' in the well is a quasi-bound positive-energy resonance associated with the shape resonance at $\lambda = 150~$nm in the partial PI cross section on the $D \rightarrow F$ channel.

\section{Discussion}


The PI cross sections warrant an experimental investigation because of the importance of optical traps of Rb and Cs Rydberg atoms in the applications mentioned near the end of Sec.~\ref{sec:Rb}, 
where atom loss and decoherence must be avoided. In our paper we stress that the model potentials used in the calculations play a central role. It is apparent that the positions of the PI minima are very sensitive to the potentials and the resultant phase shifts and quasi-bound states near and inside the Rydberg atoms' ionic cores. Noting the large depth and the small range of the inner wells of the $\ell = 3$ potentials, one may expect that a measurement of the shape resonances will present a particularly sensitive test for the $\ell=3$ model potentials.   

Considering the widths of the PI minima, one fruitful experimental approach is to use a tunable pulsed laser to photo-ionize a sample of $N$ cold Rydberg atoms and to count the ions using a single-particle counter. The latter may utilize, for instance, a micro-channel plate or a channeltron, which are capable of single-ion counting with efficiencies of $\gtrsim 30\%$. Considering that it will typically be desired to count at least one ion per laser pulse, so as to build up sufficient statistics, but fewer than $\sim N/2$ to avoid saturation, at a given $\sigma$ the fluence $F$ of the pulse should be in the range 
\begin{equation}
 \frac{h c}{2 \sigma \lambda}  \gtrsim  F \gtrsim  \frac{h c}{N \sigma \lambda}
\label{eq:f1}
\end{equation}

To measure the shape resonance of 35$D$ in Rb, this relation would have to be satisfied for $\sigma$ ranging between $\sigma_{min} \sim 200$~barn and $\sigma_{max} \sim  10000$~barn, the range of the total PI cross section of that state [see Fig.~\ref{fig:Cs_plots}~(a)]. For an assumed number of $N=10^7$ Rydberg atoms, Eq.~\ref{eq:f1} translates into
\begin{eqnarray}
    \frac{h c}{2 \sigma_{max} \lambda} \, \gtrsim \, &  F & \, \gtrsim \, \frac{h c}{N \sigma_{min} \lambda} \nonumber \\
    3 \times 10^{3} \frac{{\rm{mJ}}}{{\rm{mm}}^2} 
   \, \gtrsim \, & F & \gtrsim \, 2.7 \, \times 10^{-3} \frac{{\rm{mJ}}}{{\rm{mm}}^2} \quad . 
\label{eq:f2}
\end{eqnarray}
Noting that the PI laser could have an area of several mm$^2$, it is seen that the pulse fluence $F$ required to measure the shape resonance of Rb 35$D$ lies within fairly comfortable limits. A pulse energy of a few tens of $\mu$J per pulse could be sufficient to map out the shape resonances.

To measure the Cooper minimum of Rb 35$S$, we set 
$\sigma_{min}=\sigma_T=0.67$~barn, the elastic photon scattering rate, and $\sigma_{max} = 20$~barn. In this case, the limiting experimental requirement is 
\begin{equation}
    F \gtrsim  \frac{h c}{N \sigma_{min} \lambda} , \quad {\rm{or}}, \quad 
    F \gtrsim   0.6 \frac{{\rm{mJ}}}{{\rm{mm}}^2}   \quad . 
\label{eq:f3}
\end{equation}
It is seen that for a beam with several mm$^2$ in cross section a pulse energy of a few mJ per pulse could be sufficient to map out the Cooper minimum. This pulse energy could be delivered, for instance, by a  nanosecond pulsed dye laser, pumped with a harmonic of a pulsed YAG laser. Noting that with decreasing $n$ the PI cross sections generally increase as $n^{*-3}$, additional experimental flexibility would be afforded by lowering $n$.

A main issue with measuring the PI minima is the lasers that would be required to run these experiments. The cesium shape resonance, for example, is centered at 150~nm, which is not an easily accessible wavelength. This wavelength could be reached by running a 600~nm laser through frequency-doubling crystals, but the setup would be expensive and inefficient. Additionally, 150~nm light is readily scattered in air, so the laser beam paths must be short to avoid significant beam attenuation.

Of the five PI minima we found, we surmise that the easiest one to experimentally investigate is the Rb shape resonance (see Eq.~\ref{eq:f2}). This minimum is centered at 366~nm, which can be accessed by running a pulsed dye laser (PDL) with a dye such as LDS-720, and then sending the PDL beam through a doubling crystal. 
The Rb and Cs $S \rightarrow P$ Cooper minima are the second-best candidates for measurement because of their accessible wavelengths, but the
cross sections expected for these channels are generally low. The resultant condition on the fluence (see Eq.~\ref{eq:f3}) will make this effort more challenging.

The small total PI cross sections of Rb and Cs $nS$ Rydberg states makes these states ideal for applications of optically excitable, laser-trapped Rydberg atoms.
The Cooper minimum in Rb $\bar{\sigma}_{n, \ell=0}^{\epsilon', \ell'=1} (\lambda)$ is near $532$~nm, the second harmonic of YAG and similar lasers, which can deliver sufficient power for dipole and optical-lattice traps for Rydberg atoms~\cite{anderson11, Moore.2015a, Moore.2015b, Lampen.2018}. Laser-trapped, practically PI- and decoherence-free Rydberg atoms can be useful in applications in which atomic decay and decoherence must be minimized, such as in quantum simulation, quantum information processing and high-precision spectroscopy.


\section{Conclusion}

We have calculated the partial PI cross sections of Rb and Cs $35S$ and $35D$ Rydberg atoms from the UV into the near-IR spectral regime. We have identified one Cooper minimum and one shape resonance in Rb, and two Cooper minima and one shape resonance in Cs. 
In future work, one may investigate the PI cross sections experimentally.
The exact wavelengths of the PI minima will be useful to know for the design of optical dipole traps and optical lattices for Rydberg atoms. For instance, traps for Rb $nS$ Rydberg atoms could benefit from the Cooper minimum of Rb near $\lambda = 536$~nm. Further, the Rb shape UV lattice would be effective for Rb $nD$ Rydberg atoms. UV lattices are uncommon, but they have been used in the past to trap mercury \cite{yi11}.

Similar calculations can be performed for any element, as long as there is a model potential to use. In this vein, the same study could be conducted for other alkali metals such as potassium and sodium. While there have been articles in the past that have reported on Cooper minima in these elements, there may be other Cooper minima and shape resonances that are unknown. The studies could also be expanded out of the alkali metal group into other commonly studied species, such as Sr, Yb and Ca, which may have interesting PI behavior due to the presence of two valence electrons.
The elastic photon scattering of Rydberg atoms, which has a cross section equivalent to the Thomson scattering cross section, may also deserve a future study.

\begin{acknowledgments}

This work was supported by NSF Grant No. PHY-1707377. We thank Callum Jones of the University of California, Los Angeles, and Shruti Paranjape of the University of Michigan for valuable discussions.

\end{acknowledgments}



\bibliography{cooper}

\end{document}